\documentclass[prb,aps]{revtex4}
\usepackage{graphicx}
\usepackage{dcolumn}
\usepackage{bm}
\newcommand{\be}{\begin{equation}}
\newcommand{\ee}{\end{equation}}
\newcommand{\bea}{\begin{eqnarray}}
\newcommand{\eea}{\end{eqnarray}}
\newcommand{\beb}{\begin{eqnarray*}}
\newcommand{\eeb}{\end{eqnarray*}}
\begin{document}

\title{Fractional quantum Hall states with negative flux~:
edge modes in some Abelian and non-Abelian cases}

\author{M.~V.~Milovanovi\'{c}}
\affiliation{Institute of Physics, P. O. Box 68, 11080 Belgrade,
Serbia}

\author{Th.~Jolic\oe ur}

\affiliation{Laboratoire de Physique Th\'{e}orique et Mod\`{e}les
Statistiques, Universit\`{e} Paris-Sud, 91405 Orsay, France }

\date{\today}
\begin{abstract}
We investigate the structure of gapless edge modes propagating at the boundary
of some fractional quantum Hall states. 
We show how to deduce explicit trial wavefunctions from the knowledge
of the effective theory governing the edge modes. 
In general quantum Hall states have many edge states. Here we
discuss the case of fractions having only two such modes.
The case of spin-polarized and spin-singlet states at filling fraction $\nu =2/5$ is considered.
We give an explicit description of the decoupled charged and neutral modes.
Then we discuss the situation involving negative flux acting on the composite fermions.
This happens notably for the filling factor $\nu=2/3$ which supports
two counterpropagating modes. Microscopic wavefunctions for spin-polarized
and spin-singlet states at this filling factor are given.
Finally we present an analysis of the edge structure of a non-Abelian state
involving also negative flux.
Counterpropagating modes involve in all cases explicit derivative operators
diminishing the angular momentum of the system.
\end{abstract}
\maketitle

\section{\label{sec:level1}Introduction}

Electrons confined in a plane and subjected to a quantizing magnetic
field may display under appropriate circumstances the fractional quantum Hall effect (FQHE).
These FQHE states are gapped liquids without long-range order with unconventional properties
like fractionally charged quasiparticle excitations.
Our understanding of this phenomenon is largely based on explicit
first-quantized many-body wavefunctions.
Historically the first of these wavefunctions was introduced by Laughlin~\cite{RBL}
to describe electrons when the magnetic field is tuned so that the lowest Landau level
(LLL) has a 1/3 filling. It was soon discovered that these liquids also form
for other filling fractions, including the sequence of filling factors $\nu=p/(2p+1)$ 
which is experimentally prominent.
Some of these states can be described accurately in the framework of so-called
composite fermions~\cite{jain}. In this scheme one considers that an
even number $2p$ of fictitious flux tubes is attached to each electron,
leading to a composite object called a composite fermion (CF).
This implies that the CF now
feel a reduced flux which is equal to $B_{eff}=B-2pn\phi_0$ where
$\phi_0=hc/e$ is the flux quantum and $n$ the electron density.
Since the magnetic field is reduced, the degeneracy of the Landau levels also changes
and there are magic fillings at which an integer number of landau levels
of CFs are filled. We then expect formation of a FQHE state.
This heuristic scheme allows construction of
highly precise microscopic wavefunctions for many of the quantum
Hall states~\cite{DasSarma,Heinonen,JJBook}. It also nicely explains why
there is a compressible state at filling $\nu =1/2$ where $B_{eff} =0$
and the ground state is essentially a Fermi sea of CFs (albeit interacting).

The simple CF states however do not explain all
FQHE states observed so far. The most studied exception is the fraction $\nu =5/2$
observed in the second orbital Landau level. Due to the energy scales of the problem
it is reasonable to write this fraction as $\nu=2+1/2$ and to consider that there
is an essentially inert background of electrons fully occupying the LLL with filling $\nu=2$
and, on top of it, a half-filled landau level with interactions between electrons
renormalized by the presence of the background. So while this is again a half-filled
Landau level with zero effective magnetic field acting upon the CFs, the interaction
has been changed with respect to the LLL case. This change of interaction
is observed to destroy the CF Fermi sea which is an apt description of 
$\nu =1/2$ in the LLL and lead to formation of an incompressible  state
which is of a different kind of those already observed. This picture
assumes complete spin polarization of electrons which seems to be the case
at $\nu =5/2$. The best wavefunction
candidate to describe this new state so far is the so-called Moore-Read Pfaffian state~\cite{mr}. 
This is a microscopic wavefunction that includes some kind of pairing and
that has been constructed from correlation functions of operators
that belong to a simple two-dimensional conformal field theory.
Recent experiments have given evidence for many new fractions that do
not fit easily in existing CF scheme~\cite{Pan03,Xia04}.
These include filling factors $\nu =4/11, 5/13, 4/13, 6/17, 5/17$
and there is some weak evidence for some even denominator states
at $\nu=3/10$ and 3/8. The state at filling 3/8 is also observed in the second orbital
Landau level i.e. at $\nu=2+3/8$. When the filling factor is an odd-denominator fraction
it is plausible to argue that a hierarchical mechanism \textit{\`a la} Halperin-Haldane
is at work. For example the filling $\nu =4/11$ corresponds to an effective filling factor
1+1/3 for composite fermions and the pseudo-Landau level at filling 1/3 may
may also form a conventional Laughlin liquid. However, as in all hierarchical constructions
for fermionic constituents, there is no room for even-denominator states.

There is an interesting family of wavefunctions generalizing the Pfaffian state that are
called the Read-Rezayi states~\cite{Read96,Read99}. When written for elementary bosonic particles,
they are given by an explicit polynomial that vanishes when $k+1$ particles are at the same point
in space. For $k=1$ one finds simply the laughlin wavefunction for bosons at $\nu =1/2$,
for $k=2$ one has the (bosonic) Moore-Read Pfaffian state with filling factor $\nu =1$.
Other members of this series have $\nu =k/2$. Multiplication by an antisymmetric Jastrow
factor leads to fermionic candidate states at $\nu =k/(k+2)$. If we allow an arbitrary odd
power $M$ of the Jastrow factor, this can be extended to a family of states
at $\nu =k/(Mk+2)$. It has been noted~\cite{thj,thjTalk} that such states may be candidates even in the case
where the effective CF flux $B_{eff}$ is \textit{negative}. This leads then
to a generalized family of candidate states at $\nu =k/(3k\pm 2)$ which, surprisingly,
includes all the new fractions. There is even some limited evidence from numerical diagonalization
of small systems of electrons that these states have to do with the true ground state
of electrons in the LLL. While these new candidates are given by explicit formulas, there
are some technicalities that prevent immediate analysis. First the formulas involve
an extensive number of derivatives due to a projection onto the LLL, rendering 
analytical manipulations difficult beyond N=6 particles. Then there is no Hamiltonian
whose ground state reproduces these wavefunctions, contrary to the Read-Rezayi family.
This precludes straightforward counting of quasiparticles states or edge modes.
It is known that the effective theory describing the low energy physics of a given
quantum Hall state is encoded into the edge mode structure. For a generic
hierarchical state this edge structure is intricate and involve several fields.
however we need to understand the edge structure in order to find this effective theory.

In this paper we construct explicit wavefunctions describing the edge structure
of FQHE states involving negative flux in the CF sense. In the case of the Laughlin
state, it is well known that the one-quasihole wavefunction can be used
as the generating function for the edge modes. We use a similar construction for the
conventional fully spin-polarized CF state at filling $\nu=2/5$. The two kinds of
quasiholes leads then to two copropagating edge modes. There is also a similar picture
in the case of the spin-singlet state which can constructed also at the same filling
by couping spin-1/2 quasiholes edge modes. The simplest example of a state with negative
flux is the fraction $\nu=2/3$. While it can be considered as the particle-hole symmetric
of the Laughlin state at $\nu =1/3$, it can be viewed also as two filled pseudo Landau
levels of CF in a negative effective field. Wavefunctions constructed along this line of thought
are as successfull as those with positive flux~\cite{molsim}. We give an explicit construction
of the two counterpropagating modes from a quasihole construction. The edge modes
that propagate in the ``wrong'' direction involve derivative operators instead
of the symmetric polynomials that generate the global charge mode.
We next apply this construction to the simplest non-Abelian state with negative flux
which a Pfaffian state with bosonic filling $\nu =1$ but has a relation between
flux and number of particles different from the conventional Pfaffian state.
The edge theory is now given by a bosonic mode - the charge mode - 
and a Majorana fermion that moves in the opposite direction.

In Section II we discuss the appearance of negative flux in the CF framework.
Section III is devoted to the study of Abelian states with positive CF flux
at filling factor $\nu=2/5$, both spin-polarized and spin-singlet.
In Section IV we discuss the simplest Abelian state with negative flux, 
the fraction $\nu =2/3$. Then we apply the formalism developed in these section
to the case of the Pfaffian with positive and negative flux state in section V.
Our conclusion are given in section VI.

\section{composite fermions and negative flux}

In this section we discuss the appearance of negative flux states within the CF scheme.
We consider states of  two-dimensional electrons in the lowest Landau level (LLL).
If we use the symmetric gauge, then the one-body orbitals are given by~:
\begin{equation}
 \phi_m (z) =\frac{1}{\sqrt{2\pi m! 2^m}} {\mathrm e}^{-|z|^2/4},
\label{LLLf}
\end{equation}
where $z=x+iy$ is the complex coordinate in the plane 
and the positive integer $m$ gives the angular momentum of the state~:
$L_z =m\hbar$ (we have set the magnetic length to unity).
A general N-body LLL quantum state can thus be written as~:
\begin{equation}
 \Psi(z_1,\dots , z_N)=f(z_1,\dots ,z_N) {\mathrm e}^{-\sum_i |z_i|^2/4}.
\end{equation}
In the remainder of the paper we will always omit the (universal) exponential factor.
In an arbitrary Landau level, the one-body eigenstates now involve both $z$ and $z^*$.
A completely filled LLL state, $\nu =1$, is the Slater determinant obtained by
filling all orbitals Eq.(\ref{LLLf}) up to some maximum $m$ value. This 
(Vandermonde) determinant has the following closed form~:
\begin{equation}
\Psi_{\nu =1}(\{z_i\}) = \prod_{i<j} (z_{i} - z_{j}).
\end{equation}
This special antisymmetric product will be referred to as a Jastrow factor
in the paper. In the CF construction~\cite{jain,JJBook}, since CFs feel a reduced flux
they occupy higher Landau levels. Hence a generic CF wavefunction is~:
\begin{equation}
 \Psi_{CF}={\mathcal P}_{LLL}  \left\{ \prod_{i<j} (z_{i} - z_{j})^{2 s}
\,\, \chi^{}_{\nu^*}\right\}\, .
\label{jainr}
\end{equation} 
In this equation ${\mathcal P}_{LLL}$ is the lowest Landau level projector,
and $\prod_{i<j} (z_{i} - z_{j})^{2
 s}$, $2s$ an even integer, is the Jastrow factor that describes the flux
attachment procedure. The filling factor if the CF state is then
$1/\nu = 2s+1/\nu^*$. When we have $\nu^*=p$ an integer number of pseudo-Landau levels
then $\chi_p$ is just a Slater determinant of filled orbitals up to the $p^{th}$
Landau level. This leads to candidate wavefunctions for the prominent series of 
incompressible states at $\nu = p/(2sp+1)$. The effective magnetic field
acting on the CF is then $B_{eff}=B-2spn\phi_0$. If we fix integers $s$ and $p$ it is
clear that one can have negative flux acting upon the CFs. For example the simplest case
is $s=1$ (we are thus dealing with $^2$CFs in the notation of Jain~\cite{JJBook})
and $p=2$ i.e. at filling factor $\nu =2/3$. In the CF formalism there is nothing 
that prevents the use of wavefunctions $\chi_{\nu^*}$ with negative flux since they are simply given
by the complex conjugate of those of positive flux $\chi_{\nu^*}=\chi_{\nu}^*$. Note that in the case 
of $\nu=2/3$
there is no necessity of using the negative flux CF wavefunction since
$\nu=2/3$ can also be viewed as the particle-hole conjugate of the positive flux state at
$\nu =1/3$. In fact both approaches, negative flux or particle-hole symmetry,
give equally good wavefunctions~\cite{molsim}.

Finally we note that negative flux construction also appear in multicomponent systems.
Some convenient states with two components are those introduced~\cite{halperin} by Halperin~:
\begin{equation}
\Psi_{mm^\prime n}=
\prod_{i,j\in A}(z_i-z_j)^m
\prod_{k,l\in B}(z_k-z_l)^{m^\prime}
\prod_{a\in A,b\in B}(z_a-z_b)^n,
\label{Halperin}
\end{equation}
where the respective two-component indices belong to subsets $A$ and $B$.
This gives spin-polarized states. 
Concerning possible spin-singlet quantum Hall states, we note that
there is evidence for an incompressible state at $\nu = 2/3$ in the vanishing-Zeeman-splitting
limit~\cite{expwjd}.
Numerical studies~\cite{numwjd}
are in agreement with a spin-singlet ground state state for which the most prominent
candidate is a state~\cite{wjd} with negative flux attachment~:
\begin{equation}
\Psi_{2/3}^{(S=0)}={\mathcal P}_{LLL}\left\{ \prod_{i<j}
(z_{i\uparrow}^*-z_{j\uparrow}^{*}) \prod_{k<l}
(z_{k\downarrow}^*-z_{l\downarrow}^{*}) \prod_{p<q} (z_{p}-z_{q})^2 \right\}\, ,
\end{equation}
where $p,q$ indices run over both spin values and we have omitted the spin part
of the wavefunction.

\section{Abelian states with positive flux}

In this Section we explain how the consistent description of
the edge of negative flux Abelian states requires the inclusion of
edge states with derivative operators. Besides Abelian one-component
states we will consider spin-singlet i.e. multicomponent states for
which we will explicitly demonstrate that the existence of
derivative operators in the neutral sector is still compatible with
the charge - neutral sector separation that is expected and exists
on the edge of a fractional quantum Hall system. This will
facilitate our discussion and conclusions on the nature of
non-Abelian negative flux states, which we will consider in the
following Section using their multicomponent formulation.

\subsection{Laughlin case}
For N fully polarized fermions at filling 1/$m$ the physics of the FQHE ground state can be captured
by the Laughlin wavefunction~:
\begin{equation}
\Psi_{L}(\{z_i\}) = \prod_{i<j} (z_{i} - z_{j})^{m},
\end{equation}
where $m$ is an odd integer for antisymmetry and $i,j=1\dots N$. Above this ground state
one finds gapped excitations including charged quasiparticles.
The quasihole excitation is given by the following formula~:
\begin{equation}
\Psi_{qh}(\{z_i\};w) =  \prod_{i=1}^{N} (z_{i} - w) \Psi_{L},
\label{qh}
\end{equation}
where $w$ is the complex coordinate corresponding to the spatial location of the
quasihole. One should think of
Eq.(\ref{qh}) as a coherent state of a quasihole.
This coherent state can be expanded as a superposition of definite
angular momentum states~:
\begin{eqnarray}
\Psi_{qh}(\{z_i\};w) &=& \sum_{n = 0}^{N}  (- w)^{N -
n} s_{n}\Psi_{L}(\{z_i\}),\\
 s_{n} &=& \sum_{ i_{1} <\cdots < i_{n} } z_{i_{1}}
\cdots z_{i_{n}}.
\label{modes}
\end{eqnarray}
where $s_n$ are elementary symmetric polynomial of degree $n$. 
It is known~\cite{wenhal} that the edge states are precisely given by the products $ s_{n} \Psi_{L}; n =
1,2,\ldots$. This means that the quasihole wavefunction Eq.(\ref{qh})
can be considered
as a generating function for the edge states. 
Multiple quasihole constructions generate all combinations
(products) of symmetric polynomials, which correspond to all
possible edge states. They also emerge from the single boson effective
description of the edge of the Laughlin state~\cite{wenhal}.
Since $\Psi_{L}$ is the unique highest density zero energy
state of the hard-core interaction with interactions only for
relative angular momentum unity between electrons (for $m = 3$),
these edge states are also
zero energy states. They will smoothly transform in a low-lying manifold
of states in the presence of realistic interactions, provided the Laughlin-like
physics is preserved.

\subsection{The spin-singlet CF state at filling $\nu = 2/5$}
Next in complexity, we consider the CF state which is spin singlet for filling
$\nu=2/5$. 
We can fill the pseudo-LLL of the CFs by spin-singlet pairs only.
Since we accommodate twice as many electrons as in the polarized construction,
we obtain a global spin-singlet state at total filling $\nu=2/5$.
 If we write only the orbital part of the wavefunction it is given by~:
\begin{equation}
\label{ss2/5}
\Psi_{2/5}^{S=0}= \prod_{i<j} (z_{i\uparrow}-z_{j\uparrow})^{3}
\prod_{k<l} (z_{k\downarrow}-z_{l\downarrow})^{3}  \prod_{p<q} (z_{p
\uparrow}-z_{q \downarrow})^2.
\end{equation}
This is a multicomponent Halperin (332) state in the notation
introduced in section II. With the spin degree of freedom, it is clear
that we can now have two simple generalizations of the quasihole~:
\begin{equation}
\Psi_{qh}^{\sigma}=\prod_{i}^{N/2} (z_{i\sigma}-w) \Psi_{2/5}^{S=0},
\end{equation}
where $\sigma = \uparrow$ or $\sigma = \downarrow$. 
Each wavefunction generates a set (ring) of symmetric polynomials that
we note $s_n^{\sigma}$ defined as in Eq.(\ref{modes}). The
two sets describe the excitations of two chiral bosons on the edge
of the system.
For each pair of symmetric polynomials  of degree $n$~:
$\{s_{n}^{\uparrow},  s_{n}^{\downarrow}\}$ we can define charge
and neutral superpositions~: $\{s_{n}^{c} = s_{n}^{\uparrow} +
s_{n}^{\downarrow}, s_{n}^{s} = s_{n}^{\uparrow} -
s_{n}^{\downarrow}\} $, which are in one-to-one correspondence with two-boson 
states that describe charge and neutral excitations on the boundary
of the spin-singlet system. This construction has been introduced first by 
Balatsky and Stone~\cite{bs}.

Let us now consider the case of two quasiholes of opposite spin
at locations $w_1$ and $w_2$. The wavefunction involves a  global factor given by~:
\begin{equation}
 \Psi_{2qh} = \prod_{i} (z_{i \uparrow} - w_{1})\;
\prod_{k} (z_{k \downarrow} -
w_{2})
\times
\Psi_{2/5}^{S=0}.
\label{2qhdef}
\end{equation} 
If we expand the two-quasihole factor we find~:
\begin{eqnarray}
&&\prod_{i} (z_{i \uparrow} - w_{1})\; \prod_{k} (z_{k \downarrow} -
w_{2}) = 
\sum_{m} s_{m}^{\uparrow} w_{1}^{N/2 - m} \; \sum_{n}
s_{n}^{\downarrow} w_{2}^{N/2 - m} = \nonumber \\
=&&\frac{1}{4} \sum_{m,n} (s_{m}^{\uparrow} s_{n}^{\downarrow} +
s_{m}^{\downarrow} s_{n}^{\uparrow}) ( w_{1}^{N/2 - m} w_{2}^{N/2 -
n} + w_{2}^{N/2 - m} w_{1}^{N/2 - n})\nonumber \\
+&&\frac{1}{4} \sum_{m,n} (s_{m}^{\uparrow} s_{n}^{\downarrow} -
s_{m}^{\downarrow} s_{n}^{\uparrow}) ( w_{1}^{N/2 - m} w_{2}^{N/2 -
n} - w_{2}^{N/2 - m} w_{1}^{N/2 - n}). 
\label{decomp}
\end{eqnarray}
We thus have a  sum of two kinds of superpositions
of angular momentum eigenstates of $w$'s, each with a definite
symmetry under the coordinate exchange $w_{1} \leftrightarrow
w_{2}$. Quasiholes in the case of the spin-singlet state at $\nu =
2/5$ can be considered as spin-1/2 fermions~\cite{hal,bs}.
Therefore the first superposition is a spin-singlet $(S = 0)$,
because it is symmetric under exchange and the second
superposition is a triplet $(S = 1)$ state with $S_{z} = 0$ as it is
antisymmetric under exchange.

The important point is that the spin-singlet superposition~:
\begin{equation}
\prod_{i} (z_{i \uparrow} - w_{1})\; \prod_{k} (z_{k \downarrow} -
w_{2}) + (w_{1} \leftrightarrow w_{2}) = 
\frac{1}{2} \sum_{m,n} (s_{m}^{\uparrow} s_{n}^{\downarrow} +
s_{m}^{\downarrow} s_{n}^{\uparrow}) ( w_{1}^{N/2 - m} w_{2}^{N/2 -
n} + w_{2}^{N/2 - m} w_{1}^{N/2 - n}) 
\label{sym}
\end{equation}
generates only edge states of the charge sector exactly as a single
Laughlin (spinless) quasihole. Indeed as we take $w_{1} =
w_{2}$ in Eq.(\ref{decomp}) the coefficients in terms of $z$'s do
not change - they are the same as those that we get in the expansion
of a single spinless quasihole that generate edge states in the charge
sector. 
Therefore the family of quantities~:
\begin{equation}
S(w_{1},w_{2}) = \prod_{i} (z_{i \uparrow} - w_{1})\; \prod_{k}
(z_{k \downarrow} - w_{2}) + (w_{1} \leftrightarrow w_{2}),
\end{equation}
may be used as generators of the charge sector. In
the case of four quasiholes we can use again~:
\begin{equation}
{\mathcal S}^{(4)}(w_1,w_2,w_3,w_4)=
S(w_{1},w_{2})\; S(w_{3},w_{4}) - S(w_{1},w_{4})\; S(w_{3},w_{2}),
\label{4ss}
\end{equation}
as generators of the edge charge sector. This is
similar to the spin-singlet construction of the BCS state or in
general in the case of a many-body spin-singlet state built out of
spin-singlet pairs. Strictly speaking, in the case of the Abelian
state at $\nu = 2/5$, we do not need antisymmetrization as in
Eq.(\ref{4ss}) to generate edge states, but to keep the discussion
general and applicable to the cases that we will discuss later, we
emphasize that the charge sector can be generated through
spin-singlet $(S = 0)$ constructions of quasiholes that are made of
collections of spin-singlet pairs. 

The two kinds of states that appear
in Eq.(\ref{decomp}) are in fact orthogonal to each other. Therefore the 
associated symmetric polynomials~: $(s_{m}^{\uparrow}
s_{n}^{\downarrow} + s_{m}^{\downarrow} s_{n}^{\uparrow})$
(resp. $(s_{m}^{\uparrow} s_{n}^{\downarrow} - s_{m}^{\downarrow}
s_{n}^{\uparrow}))$ can be expressed through $s_{m}^{c} ($resp. $s_{m}^{s})$
only, because they belong to charge (neutral) sector. The 
superposition that is antisymmetric in the quasihole
coordinate exchange~:
\begin{equation}
T(w_{1},w_{2}) = \prod_{i} (z_{i \uparrow} - w_{1})\; \prod_{i}
(z_{i \downarrow} - w_{2}) - (w_{1} \leftrightarrow w_{2}),
\end{equation}
(the last line in Eq.(\ref{decomp})), creates triplet $(S = 1)$
excitations that change the spin number of the ground state and thus
generate edge states that belong to the \textit{neutral} sector. Therefore
we infer that
all possible collections of triplet pairs generate the neutral
sector. For example, in the case of four quasiholes we can use
the following combination~:
\begin{equation}
{\mathcal T}^{(4)}(w_1,w_2,w_3,w_4)=
T(w_{1},w_{2})\; T(w_{3},w_{4}) - T(w_{1},w_{4})\; T(w_{3},w_{2}),
\label{4ts}
\end{equation}
that maximizes the spin of four quasihole construction to $S_{max} =
2$.

The important conclusion is that the edge
states of the neutral sector of $\nu = 2/5$ state and, in fact, of any
two-component spin-singlet state can be generated through maximum
spin superpositions of coherent states of spin-1/2 quasiholes.

\subsection{The spin-polarized CF state at $\nu = 2/5$}

In the case of the Laughlin state the deg modes are exact zero-energy eigenstates
of the special hard-core pseudopotential for which the laughlin itself
the densest zero-energy state. But
in the case of Jain's states at $\nu = p/(2p + 1); p > 0$ 
the CF wavefunctions are not unique zero-energy ground states of special
Hamiltonians. Nevertheless we expect that the quasihole
constructions will still lead to generators of edge states. 
For the spin-polarized CF state at
$\nu = 2/5$  Jain state can be written as~:
\begin{equation}
\Psi_{2/5}= {\mathcal P}_{LLL}  \left\{ \prod_{i<j} (z_{i} - z_{j})^{2}
\,\cdot\, \chi_{2}\right\}\, ,
\label{jain}
\end{equation}
where $\chi_{2}$ represents the Slater determinant of two filled 
pseudo-Landau levels~:
\begin{displaymath}
\chi_2 =
\left| \begin{array}{cccc}
    1 & 1 & \cdots & 1 \\
    z_{1}& z_{2} & \cdots & z_{N}\\
    \vdots & \vdots& & \vdots\\
    z_{1}^{N/2} & z_{2}^{N/2} & \cdots & z_{N}^{N/2}\\
    z_{1}^{*}& z_{2}^{*} & \cdots & z_{N}^{*}\\
    z_{1}^{*} z_{1} & z_{2}^{*} z_{2}  & \cdots & z_{N}^{*} z_{N}\\
    \vdots & \vdots& & \vdots\\
    z_{1}^{*} z_{1}^{N/2} & z_{2}^{*} z_{2}^{N/2}  & \cdots & z_{N}^{*} z_{N}^{N/2}
    \end{array} \right| .
\end{displaymath}
As in the Laughlin case, we can now construct two kinds of quasiholes,
$w_{1}$ and $w_{2}$ by modifying the determinant in the following
way~:
\begin{eqnarray}
\Psi_{2qh}(w_{1},w_{2}) = \left| \begin{array}{cccc}
   (z_{1} - w_{1})  & (z_{2} - w_{1}) & \cdots & (z_{N} - w_{1}) \\
    (z_{1} - w_{1}) z_{1}& (z_{2} - w_{1}) z_{2} & \cdots & (z_{N} - w_{1}) z_{N}\\
\vdots & \vdots& & \vdots\\
(z_{1} - w_{1}) z_{1}^{N/2} & (z_{2} - w_{1}) z_{2}^{N/2} 
& \cdots & (z_{N} - w_{1})  z_{N}^{N/2}\\
    (z_{1} - w_{2}) z_{1}^{*}& (z_{2} - w_{2}) z_{2}^{*} & \cdots & (z_{N} - w_{2}) z_{N}^{*}\\
(z_{1} - w_{2}) z_{1}^{*} z_{1} & (z_{2} - w_{2}) z_{2}^{*} z_{2}  
& \cdots & (z_{N} - w_{2}) z_{N}^{*} z_{N}\\
\vdots & \vdots& & \vdots\\
(z_{1} - w_{2}) z_{1}^{*} z_{1}^{N/2} & (z_{2} - w_{2}) z_{2}^{*} z_{2}^{N/2}  
& \cdots & (z_{N} - w_{2}) z_{N}^{*} z_{N}^{N/2}
\end{array} \right|.
\label{j2qh}
\end{eqnarray}
These two kinds of quasihole corresponds to the two possibilities to create a hole
in an empty shell in the CF scheme~: we can make a hole either in the pseudo-LLL
or in the first excited orbital pseudo-LL.
When $w_{1} = w_{2}= w$ we recover the ordinary Laughlin quasihole
construction~: one can factor out $\prod_{i} (z_{i} - w)$ in front
of the ground state in Eq.(\ref{jain}).

The CF state at $\nu = 2/5$ can be viewed as a state at integer
filling factor, $\nu = 2$, of composite fermions. It is then natural
to assign  a pseudospin degree of freedom to composite fermions; those
in the pseudo-LLL  and those in the second pseudo-LL carry distinct values of
$S_{z}$, the pseudospin  number. The excitation of two quasiholes in
Eq.(\ref{j2qh}) represents two holes of composite fermions in their
respective LLs. Because the lowest energy excitations of the state
at $\nu = 2/5$ can be classified as excitations of composite
fermions in two LLs, the pseudospin $S_{z}$ quantum number can be used to
classify excitations. 
Thus, exactly as in the  case of
the spin-singlet state at $\nu=2/5$, we can consider symmetric and antisymmetric
superpositions of two quasiholes and conclude that they carry
pseudospin equal to $S = 1$ and $S = 0$
respectively. Now the presence of spin is tied to a charge excitation
and if a particular configuration of quasiholes has maximum
spin value it belongs solely to the \textit{charge} sector. 
Therefore, the symmetrized state of two quasiholes~:
\begin{eqnarray}
S^{J}(w_{1},w_{2}) &=& ( 1 + e_{12}) \Psi_{2qh}(w_{1},w_{2})
\nonumber\\
&=& \Psi_{2qh}(w_{1},w_{2}) + \Psi_{2qh}(w_{2},w_{1}),
\label{sj2qh}
\end{eqnarray}
where $e_{ij}$ denotes the exchange operation, $ i \leftrightarrow
j$, is a generator of symmetric polynomials just as a single
Laughlin quasihole. We can convince ourselves that this is true by
examining the terms in the expansion of the determinants. The
antisymmetric combination~:
\begin{eqnarray}
T^{J}(w_{1},w_{2}) &=& ( 1 - e_{12}) \Psi_{2qh}(w_{1},w_{2})
\nonumber\\
&=& \Psi_{2qh}(w_{1},w_{2}) - \Psi_{2qh}(w_{2},w_{1}),\label{jts}
\end{eqnarray}
generates the edge states of the neutral sector $(S = 0)$. These
states cannot be represented as symmetric polynomials multiplying
the Slater determinant of the ground state. The following simple
example of one electron in the LLL and one electron in second LL is
an illustration of this (we write only the determinantal part
of the wavefunction)~:
\begin{equation}
\left| \begin{array}{cc} (z_{1} - w_{1}) & (z_{2} - w_{2})\\
(z_{1} - w_{2}) z_{1}^{*} & (z_{2} - w_{2}) z_{2}^{*}
\end{array} \right| = 
z_{1} z_{2} (z_{2}^{*} - z_{1}^{*}) + w_{1} (z_{1} z_{1}^{*} -
z_{2} z_{2}^{*}) - w_{2} (z_{1} z_{2}^{*} - z_{2} z_{1}^{*}) + w_{1}
w_{2} (z_{2}^{*} - z_{1}^{*}).
\end{equation}
In the symmetric combinations of two quasiholes the second and the
third term in the expansion of the determinant combine to give $\sim
(z_{1} + z_{2})(z_{2}^{*} - z_{1}^{*})$, which demonstrates  the
factorization, a symmetric polynomial $\times$ Slater determinant in
the charge sector, which is not possible in the neutral sector~: we
get $(z_{1} - z_{2})(z_{2}^{*} + z_{1}^{*}))$.

Therefore all collections of spin-singlet pairs of quasiholes, 
generate edge states of the neutral sector of Jain's 2/5 state
We can for example use the pseudospin-singlet combination of four quasiholes~:
\begin{equation}
{\mathcal S}^{(4)}_{S=0}(w_1,w_2,w_3,w_4)=
[(1 - e_{12})(1 - e_{34}) + (1 - e_{14})(1 - e_{32})]
\Psi_{4qh}(w_{1},w_{3}; w_{2},w_{4}) \, ,
\label{j4qh}
\end{equation}
where we use the following definition~:
\begin{eqnarray}
\Psi_{4qh}(w_{1},w_{3}; w_{2},w_{4}) = \left| \begin{array}{ccc}
   (z_{1} - w_{1})  (z_{1} - w_{3}) & \cdots & (z_{N} - w_{1})(z_{N} - w_{3}) \\
    (z_{1} - w_{1}) (z_{1} - w_{3}) z_{1}& \cdots & (z_{N} - w_{1})(z_{N} - w_{3}) z_{N}\\
    \vdots & & \vdots\\
    (z_{1} - w_{1})(z_{1} - w_{3}) z_{1}^{N/2} & \cdots & (z_{N} - w_{1})(z_{N} - w_{3}) z_{N}^{N/2}\\
    (z_{1} - w_{2})(z_{1} - w_{4}) z_{1}^{*}& \cdots & (z_{N} - w_{2})(z_{N} - w_{4}) z_{N}^{*}\\
    (z_{1} - w_{2})(z_{1} - w_{4}) z_{1}^{*} z_{1} & \cdots & (z_{N} - w_{2})(z_{N} - w_{4})z_{N}^{*} z_{N}\\
    \vdots & & \vdots\\
    (z_{1} - w_{2}) (z_{1} - w_{4})z_{1}^{*} z_{1}^{N/2} & \cdots & (z_{N} - w_{2})(z_{N} - w_{4}) z_{N}^{*} z_{N}^{N/2}
    \end{array} \right|\, .
\label{j4qhdef}
\end{eqnarray}

Contrary to the previous case of neutral edge states for the  
 $\nu=2/5$ spin-singlet state, there is no factorization property of the form~:
a symmetric polynomial odd under
$\uparrow$ and $\downarrow$ exchange $\times$ the ground state.

With respect to the spin-singlet state at 2/5, we did not have a
transparent spin structure from which we could deduce edge sectors;
we used $S$ number as effectively charge number of the $U(1) \times
U(1)$ edge theory. In this case of Jain's state at 2/5, besides the
insights from the spin-singlet state, we also used the knowledge of
effective theories~\cite{read,wenhal} to argue for the existence of
two sets of generators; without going into a detailed description of
the edge states that they generate in the neutral sector, we were
able to identify them. The generators represent charge and neutral
sector, also because, as we put all quasiholes at the same point in
any generator ($w_{1} = w_{2}$ in Eq.(\ref{sj2qh}) etc.) we get a
charge hole in the charge sector, but in the neutral sector any
expression (Eq.(\ref{jts}), Eq.(\ref{j4qh}), etc.) vanishes as it
can not be connected to any charge excitation.

\section{Abelian states with negative flux}
\subsection{fully polarized CF state at $\nu = 2/3$}

We know~\cite{molsim} that the fully polarized state at $\nu
= 2/3$ can be described by both particle-hole conjugation of $\nu=1/3$ state
and negative flux CF wavefunction and thus
the edge theory should be
the same in both description.
To perform the particle-hole transformation onto the parent
Laughlin $\nu =1/3$ state one needs a background droplet with filling
 $\nu =1$ of a size larger than that of the Laughlin droplet
and one then makes the transformation in the interior region~\cite{AHM1,AHM2,wenhal}.
This leads to a inner region of filling $\nu=2/3$ separated from the vacuum
by a ring with filling $\nu=1$. As a consequence, the edge is now composite.
One can have edge modes on the exterior boundary at $\nu=1$~:
they will be generated by the symmetric polynomials we have described
in section III-A and the associated effective theory is a free boson~\cite{wenhal}.
There are also edge modes associated with the boundary between the $\nu=2/3$ core and
the $\nu=1$ annulus. These modes should propagate in the opposite direction from
the outer modes. This microscopic picture has received detailed confirmation
from numerical and experimental studies~\cite{FDM,kfp,KF,Chang,Z,DDLW,Ed}.
There are thus two counterpropagating modes that interacts probably in a non-universal manner
depending on the details of the confining potential~\cite{Ed}.
The inner modes have angular momenta which are less than the total angular
momentum of the droplet as a whole. Since its structure is exactly that of a free
boson generated by symmetric polynomials, then explicit wavefunctions for
the edge modes generators are obtained by replacing z factors by derivatives in the
generating functional Eq.(\ref{modes}). They are not exact eigenstates
but are expected to be satisfactory trial wavefunctions.


We now the discuss the edge modes as seen from the negative flux CF picture.
here we have two filled pseudo-Landau levels of $^2$CFs. A naive reasoning based
on the results for the fraction $\nu =2/5$ would lead to two copropagating modes.
However writing trial wavefunctions is in general not enough to guarantee
that their energies are as expected. For example in a general Jain state
with $p$ filled levels it is immediately clear what are the lowest-lying quasiholes
or quasielectrons since there are many possibilities for excited states.
Here we know from the reasoning of the previous paragraph that the counterpropagating
mode is generated by derivative operators~: this means that it has to be found amongst
the modes generated by the quasielectron operator~:
\begin{equation}
 {\mathcal O}_{qe}(w)=\prod_{i}(2\frac{\partial}{\partial z_i}-w^*)
\end{equation} 
Expansion in powers of $w^*$ leads to the same modes as proposed in the previous
particle-hole approach (we will not use Jain's quasielectron 
construction~\cite{jain,jeon} because it is harder to implement in the case of
many excitations and, as we consider only effective, lowest-energy
physics of edge states, the Laughlin construction is adequate).
Of course this operator does not always generate low-lying modes.
For example if applied onto the Laughlin state all modes derived from it are
gapped. For example action of $\sum_i\frac{\partial}{\partial z_i}$ will lead
to state with an excitation energy of the order of the quasielectron itself.
It is only when we act on some special states that one generates low-energy states.
Hence the operator itself does not contain all the information on the edge theory,
this is also encoded in the ground state wavefunction.


With our identification of edge modes derived from the quasielectron,
we conclude that we can use
antisymmetric combinations of pairs of  {\em quasiparticles} with pseudospin $S
=0$ to generate the neutral sector that moves in the opposite
direction of the charge sector.

%

\subsection{spin-singlet state at $\nu = 2/3$}

We now discuss the spin-singlet state $\nu =2/3$.
We know from effective theories~\cite{wenhal} that there is a chiral boson in
the edge theory that describes a neutral sector and 
moves in the opposite direction with respect to the chiral boson of
the charge sector. With the insight gained
considering the edge states for two-component spin-singlet 2/5 and
onecomponent Jain's $2/5$ and $2/3$ ground states, we conclude
that the edge states in the neutral sector of this negative flux
spin-singlet state at $2/3$ can be constructed by the action of derivative operators -
symmetric polynomials of derivatives~:
\begin{equation}
\tilde{s}_{n}^{s} = \tilde{s}_{n}^{\uparrow} -
\tilde{s}_{n}^{\downarrow},
\end{equation}
where we define~:
\begin{equation}
\tilde{s}_{n}^{\sigma} = \sum_{i = 1}^{N/2}
\frac{\partial^{n}}{\partial z_{i\sigma}^{n}},\;\;\;\; n =
1,2,3,\ldots
\end{equation}
that act on the ground state. They create excitations of the neutral
sector that move in the opposite direction from the charge
excitations. The generators of this neutral
sector are in fact the maximum spin coherent states of
spin-1/2 quasiparticles.

As emphasized in Ref.(\cite{bs}), at the edge of the $\nu = 2/5$ spin-singlet
state we have spin-charge separation and the separation is expected
in any gapped spin-singlet state. In the case
of the $\nu = 2/3$ spin-singlet state the existence of the spin-charge
separation imply that $s_{n}^{c}$ and $\tilde{s}_{m}^{s} $
commute. Indeed they do commute provided we confine our description to the
lowest-energy sector of the system i.e. the edge. This conclusion
comes from the following simple algebra~:
\begin{equation}
[s_{n}^{c}, \tilde{s}_{m}^{s}] =
[s_{n}^{\uparrow},\tilde{s}_{m}^{\uparrow}] -
[s_{n}^{\downarrow},\tilde{s}_{m}^{\downarrow}] = [s_{n}^{s},
\tilde{s}_{m}^{c}].
\end{equation}
Here by $s_{n}^{s}$ and $\tilde{s}_{m}^{c}$ we mean $ s_{n}^{s} =
s_{n}^{\uparrow} - s_{n}^{\downarrow}$ and $\tilde{s}_{m}^{c} =
\tilde{s}_{m}^{\uparrow} + \tilde{s}_{m}^{\downarrow}$. Because the
last expression does not belong to the lowest energy sector,
when the projection to the edge sector is done, we
find that $ s_{n}^{c}$ and $ \tilde{s}_{m}^{s}$ commute.


\section{Edge modes of Pfaffian and negative-flux Pfaffian states}
\subsection{The Pfaffian state in multicomponent formulation and its edge}

We now study the edge mode structure of the simplest non-Abelian quantum Hall state,
the so-called Pfaffian state introduced by Moore and Read~\cite{mr}. We consider the bosonic
case with no loss of generality since it can be multiplied by one or more odd powers
of the Jastrow factor to give an antisymmetric trial state. For bosons, its filling factor
is $\nu =1$ and the wavefunction is explicitly given by~:
\begin{equation}
\Psi_{MR}  = {\textrm{Pf}}(\frac{1}{z_{i} - z_{j}})\;\cdot\;
\prod_{i<j} (z_{i} - z_{j}),
\label{MR}
\end{equation} 
where the Pfaffian symbol stands for the following sum over permutations of N indices~:
\begin{equation}
 {\textrm{Pf}}(\frac{1}{z_{i} - z_{j}}) =\sum_{\sigma\in S_N} \textrm{sign}\sigma\,
\frac{1}{z_{\sigma (1)} - z_{\sigma (2)}}
\dots
\frac{1}{z_{\sigma (N-1)} - z_{\sigma (N)}}.
\end{equation} 
There is an alternate way~\cite{cgt} to write this wavefunction~:
\begin{equation}
\Psi_{MR} = {\mathcal S}\left\{ \prod_{i_1<j_1} (z_{i_{1}} - z_{j_{1}})^{2}
\prod_{i_2<j_2} (z_{i_{2}} - z_{j_{2}})^{2}\right\}\; ,
\label{MR-RR}
\end{equation}
where the sum is over all possible partitions of $N$ particles into
two groups denoted by 1 and 2.  It is this expression that admits a generalization
with grouping particles in $k$ subsets~\cite{Read96,Read99}.

The equivalence between these two formulas can be proved by the following manipulations~:
\begin{eqnarray}
 \Psi_{MR}=&& {\mathcal S}\left\{\prod_{i_1<j_1} (z_{i_{1}} - z_{j_{1}})^{2} \prod_{i_2<j_2}
(z_{i_{2}} - z_{j_{2}})^{2} \right\} \nonumber \\
=&&  {\mathcal S}\left\{\;\frac{\prod_{i_1<j_1}(z_{i_{1}} -
z_{j_{1}}) \prod_{i_2<j_2} (z_{i_{2}} - z_{j_{2}})}{\prod_{i_1<i_2}
(z_{i_{1}} -
z_{i_{2}})}\;\;\right\} \;\cdot \;\prod_{i<j} (z_{i} - z_{j})\nonumber \\
=&& {\mathcal S}\left\{ \textrm{Det}(\frac{1}{(z_{i_{1}} -
z_{i_{2}})})\right\} \; \cdot\; \prod_{i<j} (z_{i} - z_{j})\nonumber \\
\propto &&  \textrm{Pf} (\frac{1}{(z_{i} -
z_{j})}) \; \cdot\; \prod_{i<j} (z_{i} - z_{j}),
\label{string}
\end{eqnarray}
where we used the Cauchy determinant identity and the fact that a sum of
determinants gives a Pfaffian.

In the ``pairing'' formulation with the explicit Pfaffian Eq.(\ref{MR}), edge states are
 neutral fermion excitations  created by
breaking some of the pairs~:
\begin{equation}
\Psi_{MR}^{(n_p)}=
{\cal A}\left\{z_{1}^{m_{1}} \cdots z_{2
n_{p}}^{m_{2 n_{p}}} \frac{1}{z_{2 n_{p} + 1} - z_{2 n_{p} + 2}}
\cdots \frac{1}{z_{N - 1} - z_{N}}\right\} \;\cdot\;
\prod(z_{i} - z_{j})\, ,
\label{nfs}
\end{equation}
where $ 0 \leq m_{1} < m_{2} < \cdots < m_{2 n_{p}}$ are integers, $n_{p}$
denotes the number of broken pairs
and ${\mathcal A}$ stands for antisymmetrization as in the Pfaffian definition. 
These states can be obtained~\cite{mmnr}
 from the non-Abelian quasihole constructions~\cite{mr}for two quasiholes at 
positions $w_{1}$ and $w_{2}$~:
\begin{equation}
\Psi_{MR}^{2qh}=
\prod_{i<j} (z_{i} - z_{j})\;\cdot\; \textrm{Pf}(\frac{(z_{i} - w_{1}) (z_{j} - w_{2}) +
(w_{1} \leftrightarrow w_{2})}{z_{i} - z_{j}}).
\label{2qh}
\end{equation}
In addition to these neutral fermion modes, there is the usual charge sector
generated by symmetric polynomials as we've seen in all previous examples.

The multicomponent formulation Eq.(\ref{MR-RR}) suggests that the quasihole
factors can be introduced in 
each of the two groups of particles before symmetrizing  the whole expression.
This leads to the following alternate generating functional~:
\begin{equation}
{\mathcal G}(\tilde{w}_{1},\tilde{w}_{2})=
{\mathcal S} \left\{\prod_{i_1} (z_{i_{1}} - \tilde{w}_{1}) \prod_{i_2} (z_{i_{2}} -
\tilde{w}_{2}) \prod_{i_1<j_1} (z_{i_{1}} - z_{j_{1}})^{2} \prod_{i_2<j_2}
(z_{i_{2}} - z_{j_{2}})^{2}
\right\}\; .
\label{L2qh}
\end{equation}
One can also construct the neutral
sector from the multicomponent formulation~:
\begin{equation}
\Psi_{MR}^{(\{n_i\})}=
{\mathcal S}\left\{ s_{n_1}^{s} \cdots s_{n_k}^{s} \prod_{i_1<j_1} (z_{i_{1}} -
z_{j_{1}})^{2} \prod_{i_2<j_2} (z_{i_{2}} - z_{j_{2}})^{2}
\right\}\; ,
\label{nedge}
\end{equation}
where we have inserted
$s_{n}^{s}=s_{n}^{1} - s_{n}^{2}$. The generators - quasihole coherent
constructions of the states of the form given by Eq.(\ref{nedge})
can be easily specified in analogy with how it was done in our
discussion of the two component $\nu = 2/5$ case. 

In fact these two sets of edge states are exactly the same.
The set
Eq.(\ref{nedge}) seems to be overcomplete if we count the number of
edge states at fixed $\Delta M$ - the increase of the angular
momentum with respect to the ground state. But there are linear
dependencies among set members, Eq.(\ref{nedge}), that
will reduce their number to the number derived from
Eq.(\ref{nfs}) at fixed $\Delta M$. To show that this is true,
we come
back to the string of equivalences in Eq.(\ref{string}). Any
quasihole construction inside the Pfaffian (at the end of the
string)(like Eq.(\ref{2qh})) is also a valid quasihole construction for each
determinant in the sum over partitions. Now we have to use the
relationship between this quasihole construction and the usual
Laughlin quasihole constructions, for each determinant, explained in
Ref.(\onlinecite{mmnr}), from which it follows that the spaces of edge
states generated in these two ways are the same. This follows from the Abelian nature of
theCauchy determinant pairing. This assertion
for the Cauchy determinants automatically translates in the same
assertion for the Pfaffian~: the two kinds of quasihole constructions in
the case of the Pfaffian are different but they are superpositions of
zero energy states - edge states that belong to the same subspace of
states. Both can be used to generate this subspace and its neutral
sector. Besides Eq.(\ref{nfs}), the neutral sector can be also
represented by Eq.(\ref{nedge}) although one should remember
that this set is overcomplete.

\subsection{Negative flux Pfaffian}
We now discuss the simplest non-Abelian state with negative flux.
The basic idea is quite simple~: we start from the CF definition Eq.(\ref{jainr})
and we note that the wavefunction $\chi_{\nu^*}$ need not necessarily be a Slater determinant.
It may be itself a state with non-Abelian correlations as advocated in 
Refs.(\onlinecite{thj,thjTalk}). The simplest example is to
use a Pfaffian state for bosons. If the flux of this state is taken as positive then we simply pile up powers
of the overall jastrow factor but a more interesting possibility is to use a negative flux state
which is non-Abelian. For example the simplest example is given by
$\chi_{\nu^*=-1}=\Psi_{MR}^*$, leading to~:
\begin{equation}
\Psi_{MR}^{neg.flux}=
{\mathcal P}_{LLL} \left[{\mathcal S} \left\{\prod_{i_1<j_1} (z_{i_{1}}^{*} - z_{j_{1}}^{*})^{2}
\prod_{i_2<j_2} (z_{i_{2}}^{*} - z_{j_{2}}^{*})^{2}\right\}\; 
\prod_{i<j}(z_{i} -
z_{j})^{2}\right] .
\label{rfpf}
\end{equation}
This state has filling factor $\nu =1$ but is different from the usual Pfaffian,
for example, if written on the sphere, it has a flux-particle relationship
given by $N_{\phi}=N-1$ while the Pfaffian requires a different tuning~: $N_{\phi}=N-2$.
Not much is known about these states, some of them have interesting overlap
properties as measured in exact diagonalization of small systems~\cite{thj,thjTalk}.
In this section we construct 
the edge modes of the negative flux Pfaffian.
We  define a set of
edge states which is complete and support charge-neutral sector
separation. Our proof follows from the formalism developed in
the previous section III. We first note that
the edge states in the neutral sector can be obtained
by inserting derivative operators in the multicomponent formulation~:
\begin{equation}
{\mathcal P}_{LLL} \left[{\mathcal S} \left\{\tilde{s}_{n}^{s} \cdots \tilde{s}_{m}^{s}\prod_{i_1<j_1}
(z_{i_{1}}^{*} - z_{j_{1}}^{*})^{2} 
\prod_{i_2<j_2} (z_{i_{2}}^{*} - z_{j_{2}}^{*})^{2})\right\}\; 
\prod_{i<j} (z_{i} - z_{j})^{2}\right]\; .
\label{edgerfpf}
\end{equation}
This set is obtained by a quasiparticle coherent state construction~:
this is similar to the case of the neutral sector of the two
component $\nu=2/5$ state. 
The projection onto
the LLL, ${\mathcal P}_{LLL}$, does not induce any extra linear dependencies
among the states because the symmetrization process is analogous to
Eq.(\ref{nedge}) where every coordinate is replaced by a derivative. 
The number of edge states at fixed
$- (\Delta M)$ is then the same as the number of modes at $+\Delta M$ for 
Eq.(\ref{nedge}). Therefore Eq.(\ref{edgerfpf}), though
overcomplete, describe edge states corresponding to a massless
Majorana fermion CFT, associated to a neutral mode moving in the
opposite direction with respect to the charge mode.
We argue next that the set of modes
Eq.(\ref{edgerfpf}) allows
charge-neutral sector separation on the edge. Indeed as in the
multicomponent Abelian case, we have the following commutation~:
\begin{equation}
[s^{c}_{n},\tilde{s}_{m}^{s}] = 0.
\end{equation}
We also expect that the quantities $s_{n}^{c}$, after commuting with $\tilde{s}_{m}^{s}$'s,
act on the charge part of the ground state which is separate from
the spin part on the edge~\cite{bs}. From the multicomponent
formulation of the non-Abelian negative flux state, we can thus conclude
that the separation between modes occurs because we can
express the Pfaffian as a sum of Cauchy determinants and each of
them represents an Abelian multicomponent construction that has complete
separation between the modes.
So the conclusion is that we expect one usual charge mode and one
counterpropagating Majorana mode.

\section{Conclusions}

In this paper we have given explicit expression for some edge state wavefunctions
for fractional quantum Hall states involving more than one mode.
We discussed spin-polarized as well as spin-singlet states when there are only
two edge modes. Explicit expressions are based on the knowledge of the effective field
theory of the edge~\cite{wenhal}. Indeed if one guesses a candidate wavefunction 
it is not clear that it has to do with the edge properties. For example if we start
from the simple Laughlin state at $\nu=1/3$ and act upon the wavefunction
with derivative operators proposed for the counterpropagating mode at $\nu =2/3$
one simply generate states that have a gap of the order of the quasielectron gap.
It is only the quasihole that generate edge excitations.
So the knowledge of the ground state wavefunction is not enough, one should 
also have some knowledge of the effective theory.
We have also studied the case of negative flux states~: some of them belong
to experimentally prominent quantum Hall fractions like $\nu=2/3$. While
they were studied already by effective theory approaches~\cite{FDM,kfp,KF,Z}
 and exact diagonalizations~\cite{AHM1,AHM2,Ed}, no microscopic 
expression for the wavefunction was proposed before our work.
Finally we have studied the case of a negative flux state build upon the Moore-Read Pfaffian
whose edge modes can be constructed in a straightforward way form the formalism
we developed. The edge states are given by a bosonic charge mode and a counterpropagating
Majorana fermion. 

Of course the true electronic system is very complex due to interactions between modes.
In the case of $\nu =2/3$ it has been shown~\cite{Ed} that there is a regime
with clear separation between the two counterpropagating modes. This depends
notably of the confining potential that may reconstruct the edge. However
our explicit wavefunctions give a precise guidance to detailed numerical studies
of the edge phenomena.

\begin{acknowledgments}
We acknowledge support from a ``Pavle Savi\'c" grant and 
French-Serbian Partnership Hubert Curien. 
M. V. M. was supported by Grant No.
141035 of the Serbian Ministry of Science.

\end{acknowledgments}



\end{document}